\documentclass[10pt]{iopart}

\expandafter\let\csname equation*\endcsname\relax
\expandafter\let\csname endequation*\endcsname\relax
\usepackage{amsmath}
\usepackage{graphicx}

\begin{document}

\title{Tuning spinterface properties in Iron/Fullerene thin films}

\author{Srijani Mallik$^{1}$, Amir Syed Mohd.$^{2}$, Alexandros Koutsioubas$^{2}$, Stefan Mattauch$^{2}$, Biswarup Satpati$^{3}$, Thomas Br\"{u}ckel$^{2,4}$, Subhankar Bedanta$^{1}$}
\address{$^{1}$Laboratory for Nanomagnetism and Magnetic Materials (LNMM), School of Physical Sciences, National Institute of Science Education and Research (NISER), HBNI, Jatni-752050, India}
\address{$^{2}$J\"{u}lich Centre for Neutron Science (JCNS), Heinz Maier-Leibnitz Zentrum (MLZ), Forschungszentrum J\"{u}lich GmbH, Lichtenbergstr. 1, 85748 Garching, Germany}
\address{$^{3}$Surface Physics and Materials Science Division, Saha Institute of Nuclear Physics, 1/AF Bidhannagar, Kolkata 700 064, India}
\address{$^{4}$PGI-4: Scattering Methods Forschungszentrum J\"{u}lich GmbH 52425 J\"{u}lich, Germany}

\ead{sbedanta@niser.ac.in}
\vspace{10pt}
\begin{indented}
\item[]April 2019
\end{indented}

\begin{abstract}
In ferromagnetic (FM) metal/organic semiconductor (OSC) heterostructures charge transfer can occur which leads to induction of magnetism in the non-magnetic OSC. This phenomenon has been described by the change in the density of states in the OSC which leads to a finite magnetic moment at the OSC interface and it is called the "spinterface". One of the main motivation in this field of organic spintronics is how to control the magnetic moment in the spinterface. In this regard, there are several open questions such as (i) which combination of FM and OSC can lead to more moment at the spinterface?  (ii) Is the thickness of OSC also important? (iii) How does the spinterface moment vary with the FM thickness? (iv) Does the crystalline quality of the FM matters? (v) What is the effect of spinterface on magnetization reversal, domain structure and anisotropy? In this context, we have tried to answer the last three issues in this paper by studying Fe/C$_{60}$ bilayers of variable Fe thickness deposited on Si substrates. We find that both the induced moment and thickness of the spinterface vary proportionally with the Fe thickness. Such behavior is explained in terms of the growth quality of the Fe layer on the native oxide of the Si (100) substrate. The magnetization reversal, domain structure and anisotropy of these bilayer samples were studied and compared with their respective reference samples without having the C$_{60}$ layer. It is observed that the formation of spinterface leads to reduction in uniaxial anisotropy in Fe/C$_{60}$ on Si (100) in comparison to their reference samples.
\end{abstract}

%
\vspace{2pc}
\noindent{\it Keywords}: Spinterface, Fullerene, Polarized neutron reflectivity, Magnetization reversal, Anisotropy
%
%
%
\ioptwocol

\section{Introduction}

Organic spintronics is a transpiring field since a decade due to its potential applications in quantum computation to biomedical imaging \cite{Naber – Phys D 2007, Dediu – Nat Mater 2009, Ma’Mari – PNAS 2017}. Organic compounds are promising in hybrid metallic multilayers because of the low spin orbit coupling and hyperfine interaction associated to them \cite{Naber – Phys D 2007, Dediu – Nat Mater 2009}. This leads to long spin dephasing time ($ > 1 \mu_S$) and spin dependent transport length ($\sim$ 110 nm) which is preferable for the spintronic devices \cite{Harris – PRL 1973, Zhang – Nat Commun 2013}.   Interface plays a crucial role in such devices \textit{e.g.} controlling the spin polarization in magnetic field sensors \cite{Djeghloul – JPCL 2016}, generating spin-filtering effects in nonmagnetic electrodes \cite{Barraud – Nat Phys 2010}, modifying magnetization reversal of the parent ferromagnetic thin film \cite{Mallik – Sci Rep 2018}, altering the anisotropy symmetry \cite{Bairagi – PRL 2015}, inducing exchange bias property \cite{Gruber – Nat Mater 2015}, or giving rise to a room temperature ferromagnetism in nonmagnetic elements \cite{Ma’Mari – Nature 2015}. Charge transfer between ferromagnets (FM) and organic semiconductors (OSC) have been reported earlier where the unpaired electrons from the d orbital of the FMs are transferred to the $\pi$ orbital of the OSCs \cite{Moorsom – PRB 2014, Tran – ACS appl Mater Inter 2013, Mallik – Sci Rep 2018}. An induction of 1.2 $\mu_B$ moment per cage of C$_{60}$ and reduction of 21$\%$ moment in the Co layer at the interface was reported for Co/C$_{60}$ multilayers by polarized neutron reflectivity measurement. An antiferromagnetic coupling between the interfacial layers of cobalt and C$_{60}$ was observed from the x-ray magnetic circular dichroism measurements \cite{Moorsom – PRB 2014}. Similar hybrid interface was observed between Fe and C$_{60}$ with an induced moment of $\mu_S$ = - 0.21 and - 0.27 $\mu_B$ per molecule for C$_{60}$ prepared on Fe (001) substrate and on Fe/W (001) substrate, respectively \cite{Tran – ACS appl Mater Inter 2013, Tran – APL 2011}. Recently, we have observed that about 2 nm of C$_{60}$ close to the interface between the C$_{60}$ layer and epitaxial Fe layer on MgO (001) substrate exhibits moment $\sim$ 1.5 to 3 $\mu_B$ per cage of C$_{60}$ \cite{Mallik – Sci Rep 2018}. As mentioned above, the reason behind the charge transfer between the FM and the OSC layer is the d(FM) - $\pi$(OSC) orbital hybridization at the interface. This leads to change in the intrinsic properties of both the layers because of the modification of their density of states. When the OSC molecules are brought to contact with FM layer, depending on the geometry of the OSC molecule the density of states of the OSC gets modified by either broadening of energy levels or shifting the energy position \cite{Sanvito – Nat Phys 2010}. Further, this hybridization affects the anisotropy symmetry also.  For example it has been shown that the anisotropy can be tuned from in-plane to out-of-plane by varying the thickness of C$_{60}$ layer on Co ultrathin film \cite{Bairagi – PRL 2015}. Spin dependent hybridization has been observed between Co layer and OSC layer through Cu layer via interlayer exchange coupling \cite{Gruber – NanoLett 2015}. However, the effect of FM and OSC layer thickness on the induced moment in OSC and the magnetization reversal mechanism are not much explored.

In this paper, the magnetization reversal mechanism with real time domain images are studied by varying the thicknesses of Fe and C$_{60}$ layers in bilayer heterostructures using magneto optic Kerr effect based microscopy. The change in anisotropy due to the presence of C$_{60}$ spinterfce was evaluated using ferromagnetic resonance measurement. Further, by using polarized neutron reflectivity measurement, the induced moment in C$_{60}$ layer was quantified for variable thickness of the Fe layer.

\section{Experimental Details}

Fe/C$_{60}$ bilayer thin films of different thicknesses were prepared in a multi-deposition high vacuum chamber manufactured by Mantis Deposition Ltd., UK. The Fe and C$_{60}$ layers were prepared in-situ at room temperature using DC magnetron sputtering and thermal evaporation, respectively. The samples were grown on commercially available Si (100) substrates with a layer of native oxide (SiO$_2$). The base pressure of the system was better than 3 $\times$ 10$^{-8}$ mbar. The deposition pressure for Fe and C$_{60}$ were 5 $\times$ 10$^{-3}$ mbar and $\sim$1 $\times$ 10$^{-7}$ mbar, respectively. The rate of depositions of Fe and C$_{60}$ layers were 0.2 and 0.12 $\mathring{A}$/s, respectively. Due to the in-built geometry of our deposition system, the Fe plume was at 30$^\circ$ angle w.r.t. the substrate normal whereas the C$_{60}$ was deposited normal to the substrate \cite{Mallik,Mallik – Sci Rep 2018,Chowdhury}. To study the thickness dependence of the ferromagnetic layer on the interface between Fe and C$_{60}$, three different thicknesses of Fe and C$_{60}$ were chosen. In order to give a more understanding of the effect of thickness of the organic layer on the spinterface sample 1* has been prepared by keeping the Fe thickness same but by changing the C$_{60}$ layer thickness. To compare the change in magnetic property due to the presence of Fe/C$_{60}$ interface, three reference samples of Fe single layer of similar thicknesses were prepared. To prevent from the oxidation of Fe and damage of C$_{60}$, a capping layer of Ta was deposited in-situ in all samples using DC magnetron sputtering. The details of the samples are shown in table 1.

\begin{table*}
	\caption{Details of the prepared samples}
	\label{tbl:example}
	\begin{tabular}{ll}
		\hline
		Sample name  & Sample structure  \\
		\hline
		Sample 1   & Si (100)/SiO$_2$(native oxide)/Fe (18 nm)/C$_{60}$ (40 nm)/Ta (3 nm)   \\
		Sample 2 & Si (100)/SiO$_2$(native oxide)/Fe (7.5 nm)/C$_{60}$ (15 nm)/Ta (3 nm)  \\
		Sample 3  & Si (100)/SiO$_2$(native oxide)/Fe (3.5 nm)/C$_{60}$ (10 nm)/Ta (3 nm)  \\
		Sample 1* & Si (100)/SiO$_2$(native oxide)/Fe (18 nm)/C$_{60}$ (7 nm)	 \\	
		\hline
		Reference samples \\
		\hline
		Sample 1A   & Si (100)/SiO$_2$(native oxide)/Fe (18 nm)/Ta (3 nm)   \\
		Sample 2A & Si (100)/SiO$_2$(native oxide)/Fe (7.5 nm)/Ta (3 nm)   \\
		Sample 3A  & Si (100)/SiO$_2$(native oxide)/Fe (3.5 nm)/Ta (3 nm)   \\
		\hline
		
	\end{tabular}
\end{table*}
\normalsize

To obtain the growth of the exact layer structure of the samples cross-sectional transmission electron microscopy (XTEM) measurements has been performed on sample 1 using a high-resolution transmission electron microscope (HRTEM) (FEI, Tecnai G2 F30, S-Twin microscope, operating at 300 kV and equipped with a GATAN Orius CCD camera). The compositional analysis has also been performed by scanning transmission electron microscopy - energy dispersive x-ray spectroscopy (STEM - EDX) attachment on the Tecnai G2 F30. Energy filtered TEM (EFTEM) images were acquired using a post-column Gatan Imaging Filter (Quantum SE, model 963). The hysteresis loops and the corresponding domain images were measured simultaneously using magneto optic Kerr effect (MOKE) based microscopy manufactured by Evico magnetics GmbH, Germany\cite{Evico}. All the measurements in MOKE microscopy were performed in longitudinal mode at room temperature within a field range of $\pm$20 mT. The angle dependent hysteresis loops were measured by varying $\phi$ i.e. the angle between applied field direction and the easy axis. Polarized neutron reflectivity (PNR) at room temperature was performed on the bilayer samples (samples 1 - 3) at the MARIA reflectometer \cite{MARIA, MARIA - NEW} at at MLZ, Garching, Germany. The wavelength ($\lambda$) of the neutrons in the PNR measurements was 6.5$\mathring{A}$. The non-spin flip (NSF) scattering cross sections R$^{++}$ (up - up) and R$^{--}$ (down - down) were measured. The first and second signs in the scattering cross section correspond to the polarization of the incident and the reflected neutrons, respectively. All the PNR measurements were performed by applying the magnetic field along the easy axis of each sample at saturation, near remanence and coercivity. To evaluate the magnetic moment in single layer Fe reference samples (samples 1A - 3A), M - H loops were measured at room temperature within $\pm$500 mT using superconducting quantum interference device (SQUID) magnetometry manufactured by Quantum Design, USA. The ferromagnetic resonance (FMR) measurements were performed by Phase FMR spectrometer manufactured by NanOsc AB, Sweden\cite{FMR}. The FMR measurements were performed at a fixed frequency of 12 GHz on sample 1 and 1A to quantify the change in anisotropy. Due to low thickness and strained growth of Fe on SiO$_2$/Si (100) substrate, the FMR data measured for samples 2, 2A, 3 and 3A were very noisy. Therefore, any interpretation was not possible using FMR data of these four samples.

\section{Results and discussion}
\begin{figure*}
	\centering
	\includegraphics[width=0.9\textwidth]{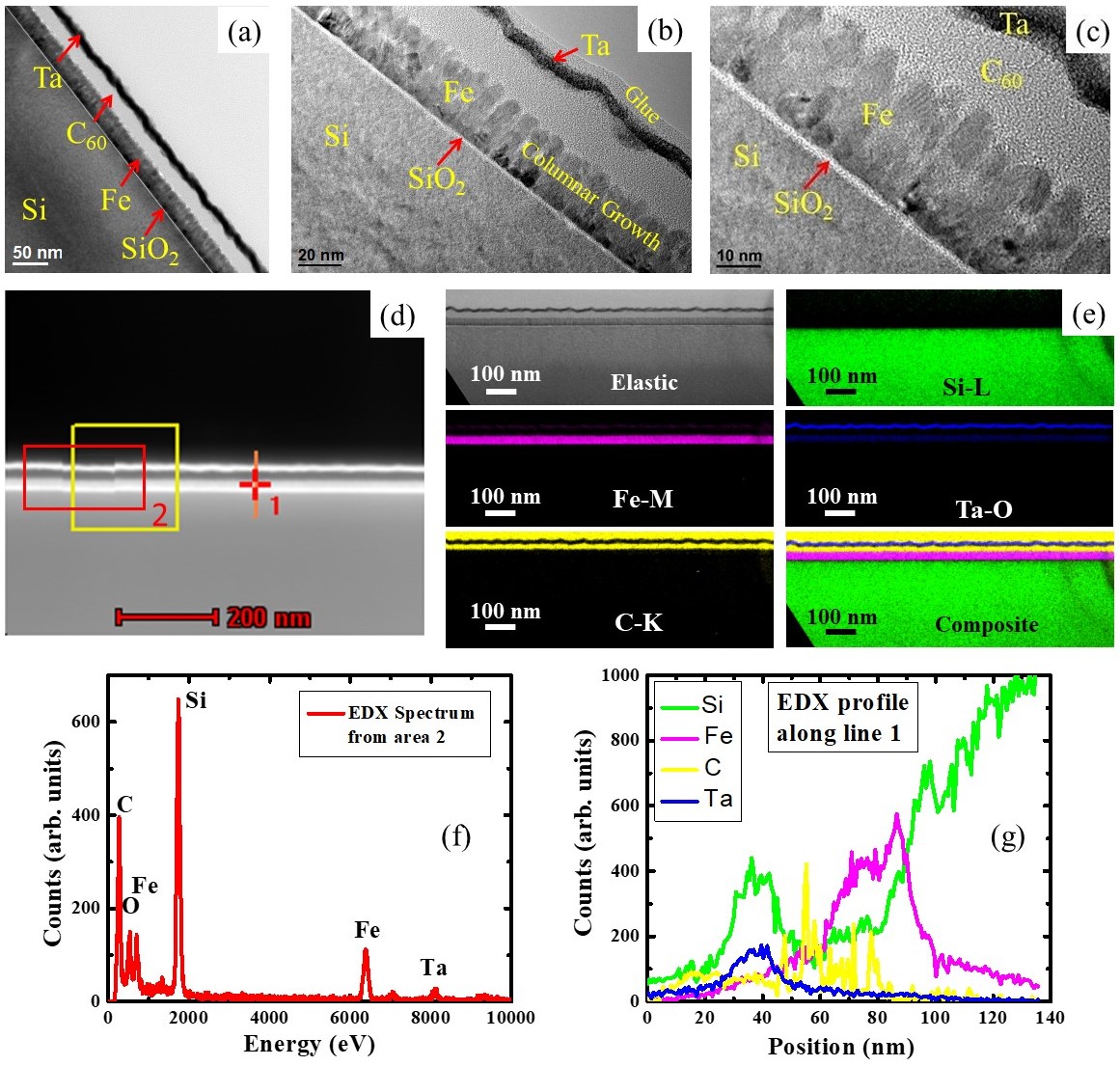}
	\caption{(a) Cross-sectional transmission electron microscopy (TEM) image of sample 1. (b) - (c) High-resolution TEM image where the columnar polycrystalline growth of Fe can be observed. (d) shows different regions of the sample where the STEM-EDX has been performed for spectrum collecting and line scan. (e) Elemental map for individual layers measured using EFTEM at different energy edges. (f) EDX spectrum showing the presence of different elements of sample 1. (g) EDX line profile to obtain the layer structure of the sample.}
	\label{fig1}
\end{figure*}

The layer structure of sample 1 has been studied using XTEM measurement. Figure 1 (a) - (c) depict the cross-sectional view of the sample 1 in different length scales. Each layer of the sample has been identified and marked in the images. It can be seen from the high-resolution TEM images (Figure 1(b) and (c)) that the next to the native oxide (SiO$_2$) layer Fe grows in columnar structure. The columnar growth of Fe can be explained from the obilque angular deposition of the Fe \cite{Bubendorff,Cherifi}. The Fe grows in polycrystalline structure due to the lattice mismatch between Fe and SiO$_2$. Different orientations of growth of Fe can be observed from the high-resolution figure 1(c). Further, STEM-EDX elemental line scan has been performed (along line 1 as shown in figure 1(d)) for each elements of different characteristic energy. EDX line profile shown in figure 1(g) confirms the layer structure of sample 1 where Ta, C and Fe peaks have been observed successively from the top surface. EDX spectrum from area 2 in (d) confirm the presence of all the different elements in the sample. The elemental mapping using EFTEM is shown in figure 1(e) where the positions of different colors represent the positions of respective layers in the sample.  Energy filtered images were acquired using Si-L and Ag-N edges and also using a contrast aperture to reduce chromatic aberrations. Chemical maps from Si L (99 eV), Fe M (54 eV), Ta O (36 eV) and C K (284 eV) edges were obtained by jump ratio method acquiring two images (one post-edge and one pre-edge), respectively, to extract the background, with an energy slit of 8 eV for Si, 4 eV for Fe and Ta and 10 eV for C, respectively. The elastic (zero loss) and composite images are shown in Figure 1(e). It should be noted that the glue used in the sample preparation of cross-sectional TEM measurement contains C. Therefore, signal for C has been detected on top the Ta layer also.  

The layer specific magnetic moments and interfacial magnetic properties were studied using PNR technique in specular reflectivity mode for all the bilayer samples. The incident polarized neutrons interact with the magnetic spins of the sample and flip their sign according to the sign of the interacting spin. The intensity of the reflected neutrons (both up and down) was measured as a function of the component of the momentum transfer (Q$_Z$) which is perpendicular to the sample surface. The relation between Q$_Z$ and the incidence/reflected angle is Q$_Z$ = $\frac{4\pi}{\lambda}sin\theta$ where $\theta$ is the angle of incidence/reflection and $\lambda$ is the neutron wavelength. It is possible to acquire layer specific magnetic information of a multilayer sample by selecting a suitable scan range of Q$_Z$ as it is a variable conjugate to the depth \textit{d} from the surface of the film\cite{Paul}.

\begin{figure}[h!]
	\centering
	\includegraphics[width=0.5\textwidth]{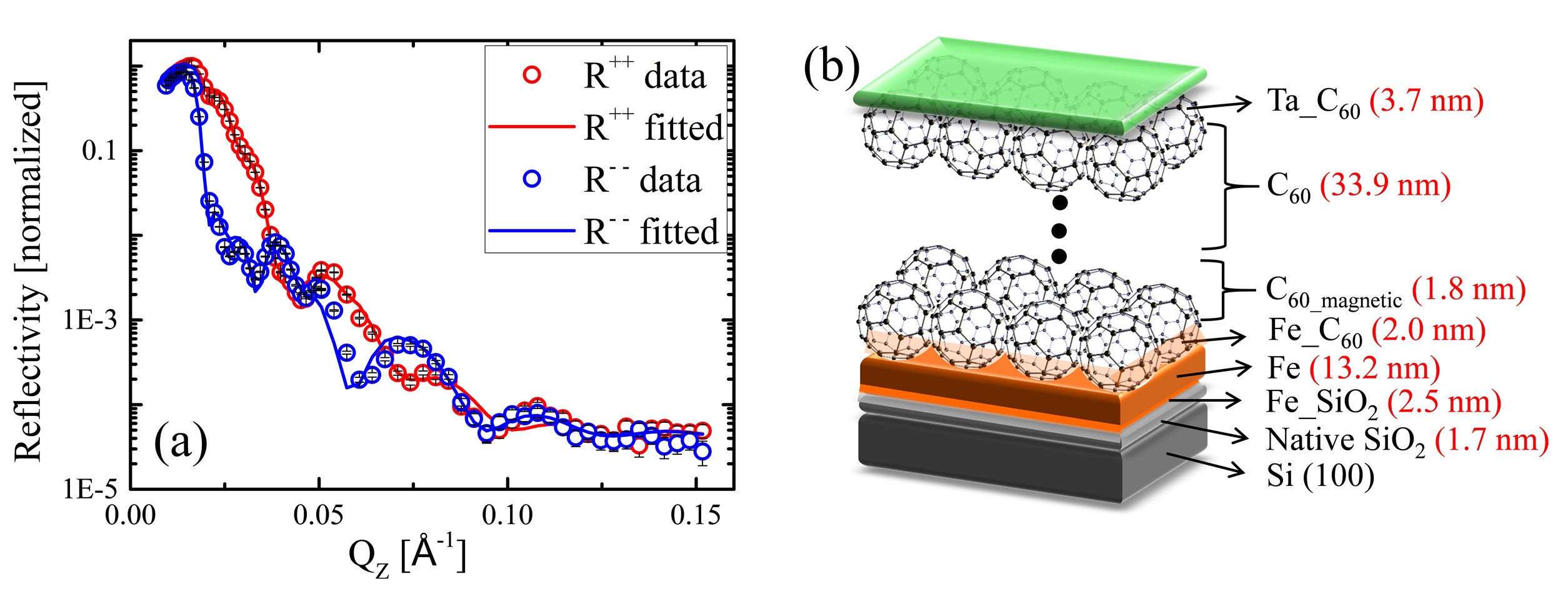}
	\caption{(a) Polarized neutron reflectivity data and the corresponding fits for sample 1. The red and blue open circles represent the data measured for the up - up and down - down channels, respectively. The solid lines correspond to their respective fits. The measurement was performed at the saturation state at room temperature. (b) A schematic representation of the sample structure obtained by fitting the PNR data shown in (a). The numbers written in brackets beside each layers correspond to the fitted thickness of the respective layer.}
	\label{fig2}
\end{figure}

Figure 2(a) shows the PNR data (open circles) and the corresponding fits (solid lines) for sample 1 measured at room temperature at the positive saturation state ($\mu_0H$ = 100 mT) of the sample. The reflectivity for the non-spin flip (NSF) scattering cross section (i.e. R$^{++}$ and R$^{--}$) was recorded where the red and blue open circles represent the data measured for the up - up and down - down channels, respectively (Figure 2(a)). The fitting was performed using GenX software \cite{GENX} which is based on the Parratt formalism \cite{Parratt}. The fits reveal the presence of interdiffusion in between all the layers of sample 1. The schematic sample structure obtained from the fitting is depicted in figure 2(b) where the thicknesses are mentioned beside the respective layers. From the PNR data analysis the magnetic moment of the Fe layer in sample 1 is obtained to be 1.56 $\pm$ 0.06 $\mu_B$/atom. For the sake of comparison the M-H loop for the reference sample 1A was measured using SQUID magnetometer (Figure S1 of the supplementary information) and the magnetic moment is calculated to be 1.98 $\pm$ 0.05 $\mu_B$/atom. Therefore, a loss of $\sim$21$\%$ of magnetic moment is observed for sample 1 with respect to its reference sample 1A. An interdiffusion layer of 2 $\pm$ 0.05 nm is observed between the Fe and C$_{60}$ layer. Next to this mixed layer, 1.8 $\pm$ 0.06 nm of pure C$_{60}$ layer exhibits a magnetic moment of 1.9 $\pm$ 0.45 $\mu_B$/cage of C$_{60}$. The interdiffusion layer shows a magnetic moment of 2.4 $\pm$ 0.6 $\mu_B$ per unit where 1 unit consists of one Fe atom and one C$_{60}$ cage. The moment in the interdiffusion layer is higher as it contains moment from both the Fe and magnetic C$_{60}$. The reason behind the loss of moment in Fe and induction of moment in C$_{60}$ cage can be ascribed to the hybridization of d and $\pi$ orbitals between the Fe and C, respectively. Due to the electronic configuration of C atoms the C$_{60}$ cages have affinity for electrons and on the other hand Fe can donate extra unpaired electron from its d orbital. Due to this $\pi$ - d hybridization, the density of states of the C$_{60}$ get modified which leads to ferromagnetism in fullerene \cite{Sanvito – Nat Phys 2010}. However, the thickness of charge/spin transfer from Fe layer to C$_{60}$ layer is limited to 1.8 $\pm$ 0.06 nm. Next to the magnetic C$_{60}$ layer, remaining 33.9 $\pm$ 0.03 nm of C$_{60}$ layer does not exhibit any signature of ferromagnetism. This sample was also measured near remanence ($\sim$2 mT) to elucidate the magnetization direction of the layers near zero field (Figure S2(a) in supplementary information). It has been observed from the best fit that 80$\%$ of the Fe layer spins point in negative direction whereas the magnetic C$_{60}$ layer shows positive moment. This indicates that the Fe and magnetic C$_{60}$ layers are anti-parallel to each other near remanence. This result is in good agreement with our recent report on Fe/C$_{60}$ bilayer grown on MgO (001) substrate \cite{Mallik – Sci Rep 2018}. In the latter case, Fe layer was grown epitaxially on MgO (001) substrate and C$_{60}$ exhibits $\sim$2.95 $\mu_B$/cage of moment \cite{Mallik – Sci Rep 2018}. However, due to the polycrystalline growth of Fe on native oxide based Si (100) substrate, the induced moment in C$_{60}$ is less in this case. Therefore, it seems that the crystalline quality of the ferromagnetic layer probably has consequences on the strength of the induced moment in the C$_{60}$ layer. In order to understand the effect of thickness of the organic layer on the spinterface, PNR measurement has been performed on sample 1*. The fitted data and the sample structure obtained from the PNR fit is shown in figure S3. It has been observed that the thickness of non-magnetic C$_{60}$ does not contribute to the spinterface properties.

\begin{figure}[h!]
	\centering
	\includegraphics[width=0.5\textwidth]{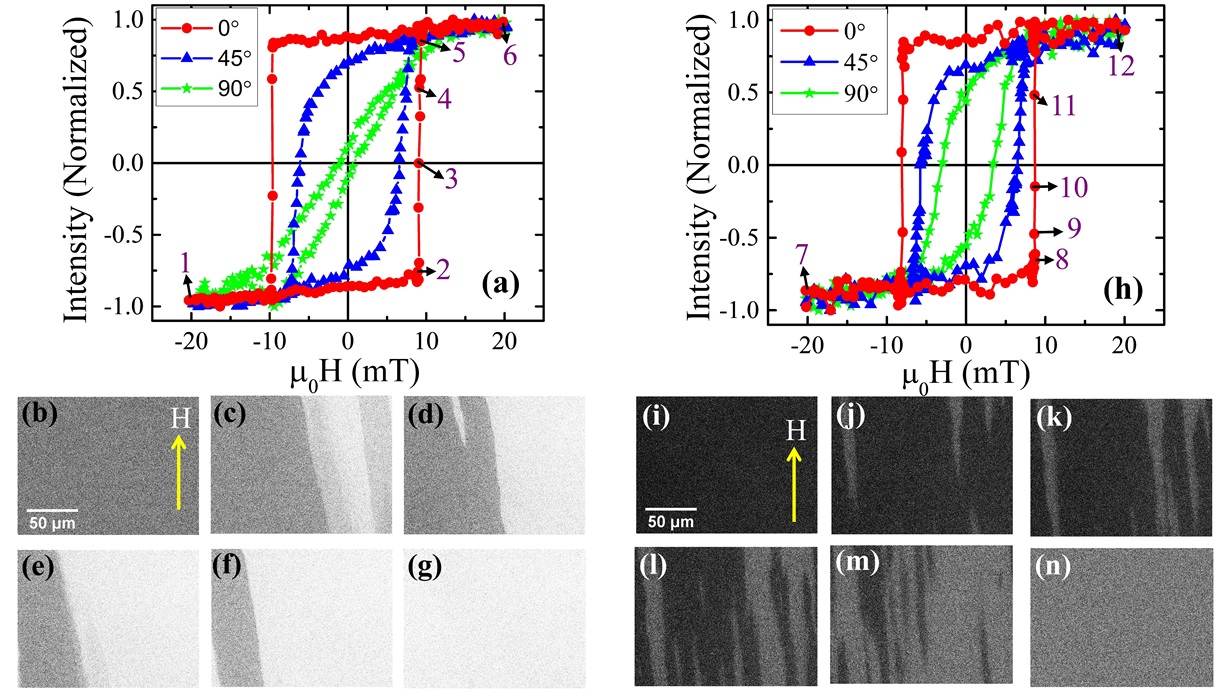}
	\caption{Hysteresis loops measured using magneto optic Kerr effect based microscopy in longitudinal mode by varying the angle ($\phi$) between the applied field and the easy axis at room temperature for samples (a) 1A and (h) 1, respectively. The images shown in (b) - (g) correspond to the domain states observed in sample 1A along the easy axis (red curve with solid circles) at the field points 1 - 6, respectively. Similarly, the images shown in (i) - (n) correspond to the domain states observed in sample 1 along the easy axis (red curve with solid circles) at the field points 7 - 12, respectively. All the domain images are in same length scale shown in (b) and (i). The arrows shown in image (b) and (i) represent the direction of the applied field.}
	\label{fig3}
\end{figure}

To study the effect of this magnetic C$_{60}$ layer on the magnetization reversal of the Fe layer, angle dependent hysteresis loops with simultaneous domain imaging were performed on samples 1 and 1A using MOKE microscopy in longitudinal mode at room temperature. Figure 3(a) shows the hysteresis loops measured along $\phi$ = 0$^\circ$ (easy axis), 45$^\circ$ and 90$^\circ$ for sample 1A. It can be observed from the coercivity variation of the hysteresis loops shown in figure 3(a) that sample 1A exhibits uniaxial anisotropy. Due to the geometry of our deposition chamber, Fe plume was at an angle of 30$^\circ$ with respect to the substrate normal. Due to the oblique angle of deposition, uniaxial anisotropy is induced in the system \cite{Cherifi,Bubendorff,Mallik,Chowdhury,Chowdhury - TSF, Mallik - JMMM1, Mallik - JMMM2}. It has been discussed in previous reports that the grains of the material form chain like structure along the perpendicular to the in-plane projection of the plume direction. Along this direction, an elongation in the grain structure is expected which in turn induces a uniaxial anisotropy in the system \cite{Cherifi,Bubendorff,Mallik,Chowdhury,Chowdhury - TSF, Mallik - JMMM1, Mallik - JMMM2}. The origin of the oblique angle of deposition induced uniaxial anisotropy is the long range dipolar interaction between the grains \cite{Cherifi,Bubendorff,Mallik}. The domain images shown in figure 3 (b) - (g) correspond to the domain states observed along the easy axis at points 1 - 6, respectively, for sample 1A. Large stripe domains are observed in this sample. It is noted that Fe exhibits polycrystalline growth on Si (100) substrate with native oxide layer due to the large lattice mismatch between the lattice constant of bcc Fe (0.278 nm) and hcp SiO$_2$ (a = 0.491 nm; c = 0.54 nm). This induces strain in the Fe layers which may add dispersion in the oblique angle deposition induced uniaxial anisotropy direction. Stripe domains are usually observed in perpendicularly magnetized films as a net result of competition between the anisotropy and demagnetization energy \cite{Viret}. However, in-plane magnetized films can also exhibit stripe domains due to the effect of strain arising due to the inhomogeneity leading to a dispersion in uniaxial anisotropy \cite{Hubert}. It has also been reported that stripe domains are observed for obliquely grown materials due to their columnar growth (Figure 1(b)) \cite{Hubert}. Figure 3(h) shows the hysteresis loops for sample 1 measured along $\phi$ = 0$^\circ$, 45$^\circ$ and 90$^\circ$. Sample 1 also exhibits uniaxial anisotropy. However, the difference of the coercivity between easy and hard axis is less for the Fe/C$_{60}$ bilayer sample in comparison to its corresponding reference sample 1A. This indicates reduction in the strength of the uniaxial anisotropy in sample 1. Due to the amorphous growth of C$_{60}$ on polycrystalline Fe, misalignment is probable between neighbouring grains which increases the dispersion in the anisotropy. This leads to decrease in the size of the stripe domains (Figure 3(i) - (n)). As the dispersion in the net anisotropy increases in sample 1 the formation of the stripe domains become denser in comparison to the reference sample 1A \cite{Hubert}.

\begin{figure}
	\centering
	\includegraphics[width=0.5\textwidth]{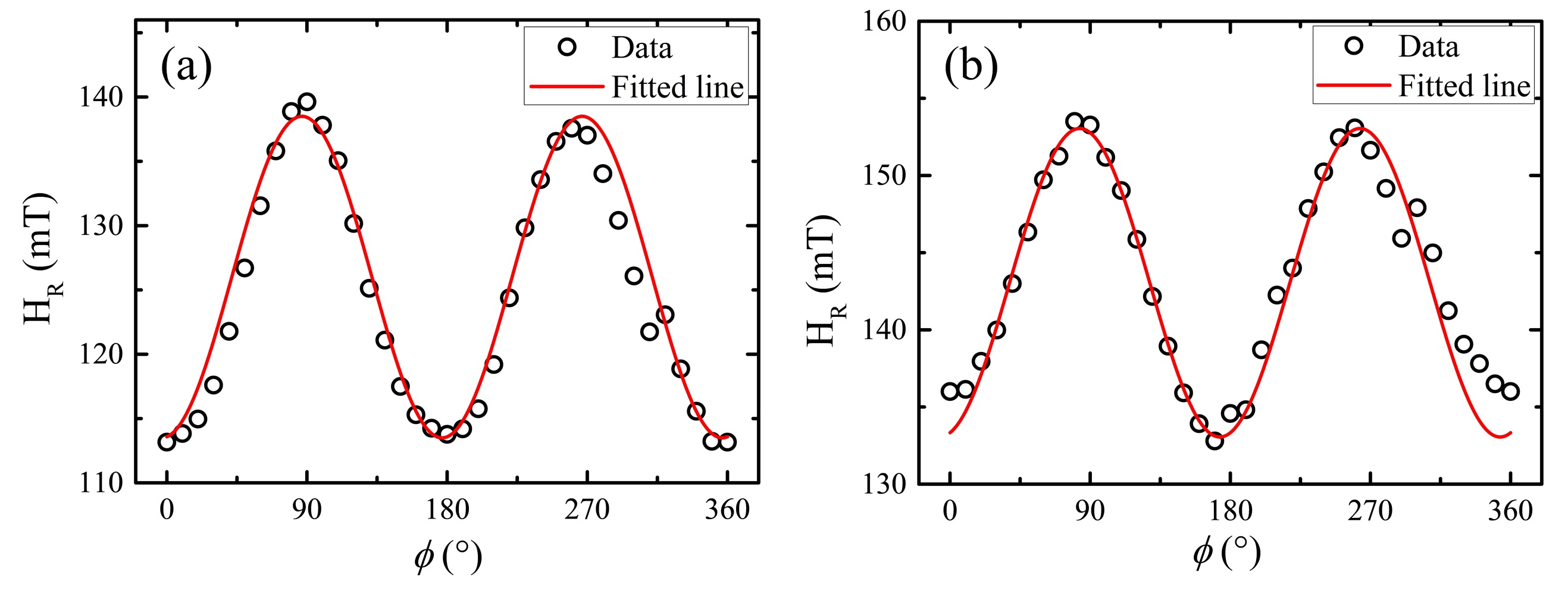}
	\caption{Anisotropy symmetry plot and the fits for (a) sample 1A and (b) sample 1 measured using ferromagnetic resonance (FMR) technique by keeping the frequency fixed at 12 GHz. The black colored open circles and the red line correspond to the measured data and the fit to equation 2, respectively.}
	\label{fig4}
\end{figure}

To quantify the change in anisotropy due to the presence of magnetic C$_{60}$ interface, angle dependent FMR measurement has been performed on samples 1 and 1A at a fixed frequency of 12 GHz. The angle $\phi$ between the easy axis and the applied field direction have been varied with an increment of 10$^\circ$ in the full 360$^\circ$ sample rotation. The resonance field $H_R$ has been recorded for each angle and the anisotropy nature of the sample has been extracted by plotting $H_R$ as a function of $\phi$ (Figure 4). The anisotropy constant of the samples can be derived by fitting these data in terms of the energies present in the system. The total magnetic free energy density of the system can be written as \cite{Gieniusz - JMMM 2007}: 
\begin{equation}
\begin{aligned}
E = & - HM_S[sin\theta sin\theta_M cos(\phi_M - \phi) + cos\theta cos\theta_M] \\
 & - 2\pi M_S^2 sin^2\theta_M + K_Usin^2\theta_M + K_{in}sin^2\theta_M sin^2\phi_M\label{eqn:example}
\end{aligned}
\end{equation}
where $K_{in}$ is the in-plane uniaxial anisotropy constant and $K_U$ is the perpendicular contribution of the anisotropy. $\phi$ is the angle between the easy axis and the projection of the applied field direction in the sample plane. $\phi_M$ is the angle between the easy axis and the projection of magnetization in sample plane. $\theta$ and $\theta_M$ are the angles between the z-axis wrt the applied field direction and the magnetization direction, respectively. However, our samples are in-plane magnetized and the external field is also applied in the sample plane. Therefore, $\theta_M$ and $\theta$ are considered to be 90$^\circ$. To evaluate the strength of the anisotropies present in samples 1 and 1A, angle dependent $H_R$ was fitted using the following dispersion relation \cite{Gieniusz - JMMM 2007}
\begin{equation}
\begin{aligned}
(\frac{\omega}{\gamma})^2 = & [Hcos(\phi_M - \phi) - h_U + h_{in}sin^2\phi_M][Hcos(\phi_M - \phi) \\
& - h_{in} + 2h_{in}sin^2\phi_M]\label{eqn:example}
\end{aligned}
\end{equation}
where, $h_{in}$ is the in-plane anisotropy representations and can be written as $h_{in}$ = $\frac{2K_{in}}{M_s}$ ($K_{in}$ is the in-plane uniaxial anisotropy constant). $h_U$ is the out-of-plane anisotropy representation and considered to be negligible as the samples are in-plane magnetized. The red solid lines in figure 4 show the fitted curves for the angle dependent $H_R$ data using equation 2 for both the samples. The anisotropy field $h_{in}$ is extracted to be 25 and 20 mT for samples 1A and 1, respectively. The value of saturation magnetization ($M_S$) for both the samples was measured using SQUID magnetometer. The uniaxial anisotropy constant $K_{in}$ is calculated to be 2.16 $\times 10^4$ and 1.05 $\times 10^4 J/m^3$ for samples 1A and 1, respectively. Therefore, it can be concluded that the anisotropy decreases by $\sim 51\%$ for sample 1 than that of its corresponding reference sample 1A. Hence, it is believed that the dispersion in the anisotropy in sample 1, due to the lattice mismatch between C$_{60}$, Fe, SiO$_2$, and Si (100) substrate, is the reason behind the decrease in net uniaxial anisotropy of the system.

\begin{figure}
	\centering
	\includegraphics[width=0.5\textwidth]{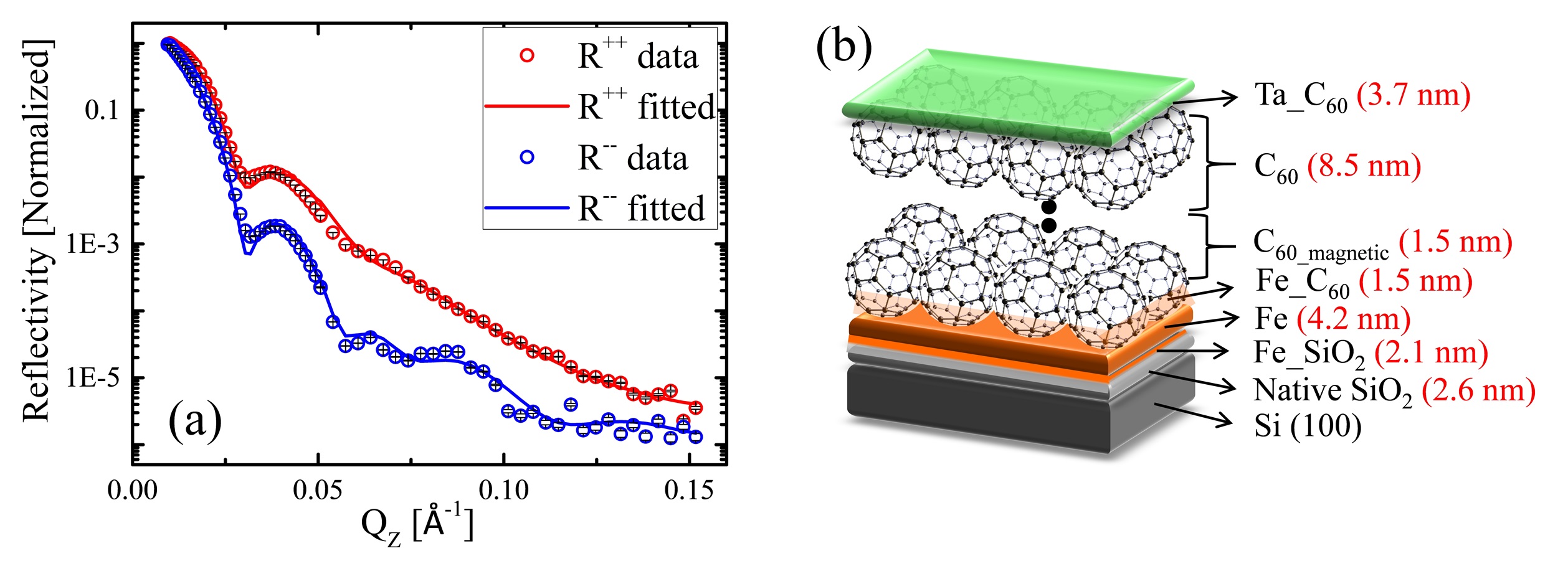}
	\caption{(a) Polarized neutron reflectivity data for sample 2. The red and blue open circles represent the data measured for the up - up and down - down channels, respectively. The solid lines correspond to their respective fits. The measurement was performed at the saturation state at room temperature. (b) A schematic representation of the sample structure obtained by fitting the PNR data shown in (a). The numbers written in brackets beside each layers correspond to the fitted thickness of the respective layers.}
	\label{fig5}
\end{figure}

\begin{figure}
	\centering
	\includegraphics[width=0.5\textwidth]{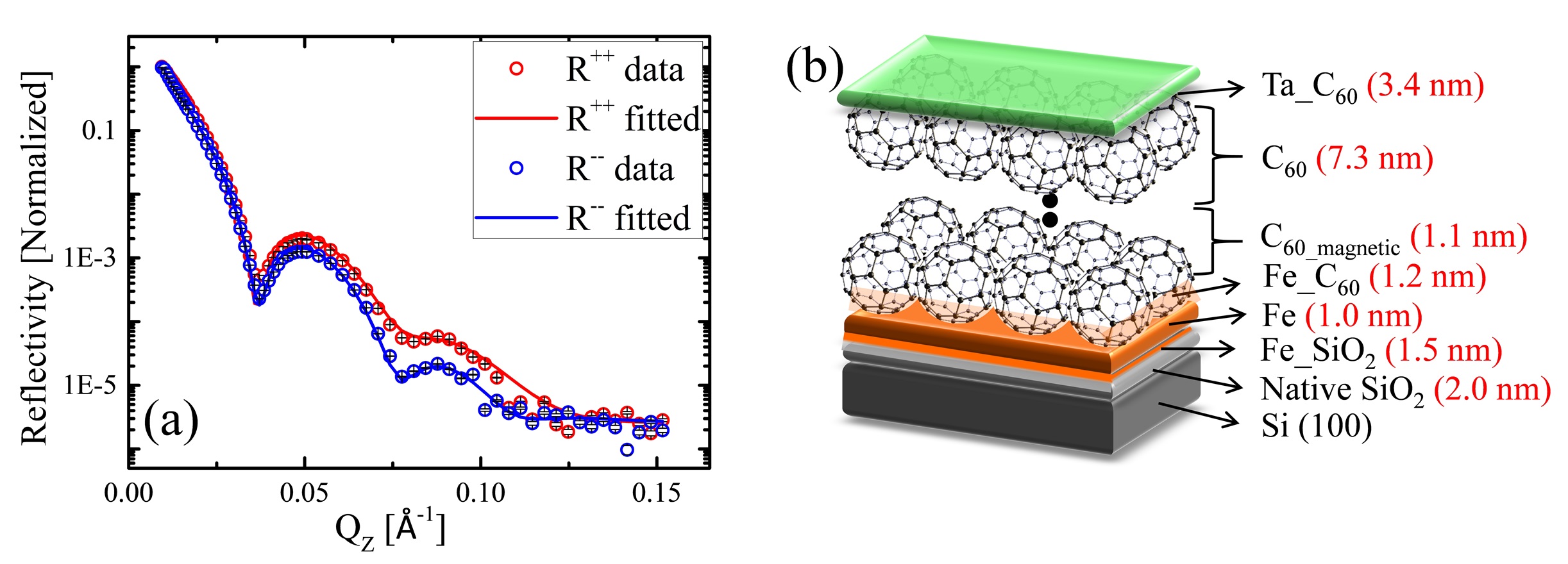}
	\caption{(a) Polarized neutron reflectivity (PNR) data for sample 3. The red and blue open circles represent the data measured for the up - up and down - down channels, respectively. The solid lines correspond to their respective fits. The measurement was performed at the saturation state at room temperature. (b) A schematic representation of the sample structure obtained by fitting the PNR data shown in (a). The numbers written in brackets beside each layers correspond to the fitted thickness of the respective layers.}
	\label{fig6}
\end{figure}

To understand the effect of the Fe layer thickness on the induced moment in the C$_{60}$ layer, PNR measurements have been performed on samples 2 and 3. The PNR data and their corresponding fits at the saturation state for samples 2 and 3 are shown in figure 5(a) and 6(a), respectively. Sample 2 exhibits a spinterface of 1.5 $\pm$ 0.07 nm of C$_{60}$ (Figure 5(b)). The induced moment in C$_{60}$ is extracted to be 1.52 $\pm$ 0.5 $\mu_B$/cage. Therefore, sample 2 shows less induced moment than that of sample 1. The moment in Fe layer was calculated from SQUID and PNR measurement for samples 2A and 2, respectively. The Fe layer exhibits 1.77 $\pm$ 0.04 $\mu_B$/atom (Figure S1(b) in supplementary information) and 1.36 $\pm$ 0.08 $\mu_B$/atom in samples 2A and 2, respectively. The loss in Fe moment is $\sim$23$\%$ for sample 2 in comparison to its corresponding reference sample 2A. The PNR data and its fit for sample 2 near the remanence state are shown in figure S2(b) in the supplementary information. Similar to the sample 1, the Fe and C$_{60}$ layers exhibit anti-parallel coupling at the remanance state. The Fe layer is 87$\%$ reversed near the remanence whereas the C$_{60}$ layer is still in the positive state. By fitting the saturation data for sample 3, it has been observed that only 1.37 $\pm$ 0.18 $\mu_B$/cage of magnetic moment is induced in C$_{60}$ at the interface. The thickness of the magnetic C$_{60}$ is 1.1 $\pm$ 0.08 nm (Figure 6(b)) which is the thickness of a monolayer of C$_{60}$ \cite{Qiao - NanoLett}. This indicates that the induced moment in C$_{60}$ decreases monotonically with the decrease in Fe thickness. However, the loss of Fe moment is highest in sample 3. The Fe moment in its reference sample 3A have been obtained to be 1.49 $\pm$ 0.08 $\mu_B$/atom by SQUID magnetometry (shown in supplementary figure S1(c)). Whereas in sample 3 the moment in Fe is observed to be only 1.06 $\pm$ 0.02 $\mu_B$/atom which corresponds $\sim$29$\%$ loss in the Fe moment. The growth of Fe on Si (100) substrate with native oxide is poor in the ultrathin limit as the roughness of the Fe layer becomes nearly in the order of its thickness. Also, the interdiffusion between the Fe and C$_{60}$ layer is even more than the pure Fe layer thickness (Figure 6(b)). Due to the presence of such disorder in the growth of the Fe, the moment decreases rapidly. The PNR data and fits near the remanence state for sample 3 is shown in figure S2(c) of the supplementary information. Here, the Fe layer shows 76$\%$ negative moment and the C$_{60}$ layer is completely in its positive state. Therefore, it can be concluded that irrespective of the Fe layer thickness, the magnetic C$_{60}$ and the Fe layer shows anti-parallel magnetization at the remanence state.
It has been observed from the PNR fittings of samples 1 - 3 that both the moment and the thickness of the spinterface increases with increase in the Fe layer thickness. This can be explained in terms of the quality of growth of the Fe layer. The lattice constant of bcc Fe and hcp SiO$_2$ is 0.278 and 0.491 nm, respectively. Due to this huge lattice mismatch, additional strain might be induced in the Fe layers. In ultrathin films the strain felt by the Fe is even more as they do not have sufficient thickness to relax the adatoms. It is expected that the growth of C$_{60}$ will be also poor on such strained Fe surface. Therefore, the induced moment in C$_{60}$ is less when grown on ultrathin Fe. But for thicker Fe samples the topmost Fe adatoms are much relaxed in comparison to the ones next to the Fe/SiO$_2$ interface.  Therefore, the induced moment in C$_{60}$ increases with the increase in Fe layer thickness. Similarly, the thickness of the spinterface is greater for the samples having thicker Fe layers. The magnetization reversal mechanism, domain structure and anisotropy symmetry of samples 3 and 3A are shown in figure S4 in the supplementary section. Similar to the thicker samples, the domain structure becomes smaller for sample 3 due to the presence of spinterface. Also, the anisotropy decreases in sample 3 when compared to its reference sample 3A. This is in good agreement with our previously obtained behavior in the thicker sample 1.

\section{Conclusion}

Fe/C$_{60}$ bilayer samples have been prepared on Si(100) substrate with native oxide by varying the thickness of Fe layer. Formation of spinterface have been observed in all the samples due to the $\pi$ - d hybridization between C and Fe atoms. The induced moment in C$_{60}$ is highest for the thicker (t$_{Fe}$ = 18 nm) sample. Due to the lattice mismatch between Fe and SiO$_2$, the growth of Fe is polycrystalline in nature. Because of this growth induced strain in Fe, both the moment and thickness of the spinterface decreases with decrease in the Fe layer thickness. The formation of spinterface leads to decrease in uniaxial anisotropy in comparison to their reference samples without having the C$_{60}$ layer. Such studies may help in choosing right parameters for fabricating TMR devices which has potential application in the organic spintronics.

\section*{Acknowledgments}
The authors thank Department of Atomic Energy, and Department of Science and Technology - Science and Engineering Research Board (EMR/2016/007725), Govt. of India, for providing the financial support to carry out the experiments. The authors also thank the Department of Science and Technology, India (SR/NM/Z-07/2015) for the financial support for performing the neutron experiments and Jawaharlal Nehru Centre for Advanced Scientific Research (JNCASR) for managing the project.

\end{document}